\documentclass{iau} 
\usepackage{graphicx}
\usepackage{natbib}
\usepackage{hyperref}

\title[Magnetic survey of classical Cepheids] 
{First results of a magnetic survey of classical Cepheids}

\author[James A. Barron, Gregg A. Wade, Colin P. Folsom \& Oleg Kochukhov]   
{James A. Barron$^{1,2}$,
 Gregg A. Wade$^2$, Colin P. Folsom$^3$ \\ \and Oleg Kochukhov$^4$}

\affiliation{$^1$Department of Physics, Engineering Physics \& Astronomy, Queen’s University, Canada \\ email: {\tt j.barron@queensu.ca} \\[\affilskip]
$^2$Department of Physics and Space Science, Royal Military College of Canada\\[\affilskip]
$^3$Tartu Observatory, University of Tartu, Estonia
\\[\affilskip]
$^4$Department of Physics and Astronomy, Uppsala University, Sweden}

\pubyear{2022}
\volume{361}  
\jname{Massive Stars Near and Far}
\editors{N.~St-Louis, J.~S.~Vink \& J.~Mackey, eds.}
\begin{document}

\maketitle

\begin{abstract}
We report recent ESPaDOnS and HARPSpol spectropolarimetric observations from our ongoing magnetic survey of the brightest twenty-five classical Cepheids. Stokes~$V$ magnetic signatures are detected in eight of fifteen targets observed to date. The Stokes~$V$ profiles show a diversity of morphologies with weak associated longitudinal field measurements of order 1~G. Many of the Stokes $V$ profiles are difficult to interpret in the context of the normal Zeeman effect. They consist of approximately unipolar single or double lobe(s) of positive or negative circular polarization. We hypothesize that these unusual signatures are due to the Zeeman effect modified by atmospheric velocity or magnetic field gradients. In contrast, the Stokes~$V$ profiles of Polaris and MY~Pup appear qualitatively similar to the complex magnetic signatures of non-pulsating cool supergiants, possibly due to the low pulsation amplitudes of these two stars.
\end{abstract}

\firstsection 
\section{Introduction}
Classical Cepheid variables are essential tools in studying stellar evolution and cosmology due to their radial pulsations and period-luminosity relation (e.g. \citealt{riess_2022}). As luminous yellow supergiants, Cepheids are the evolutionary descendants of main-sequence intermediate to massive B-type stars. While Cepheids have been studied for over a century, we know surprisingly little about how magnetic fields may influence their evolution and stellar properties.

\begin{figure}[h!]
\begin{center}
 \includegraphics[trim={0 1.1cm 0 1.6cm},clip,width=0.87\textwidth]{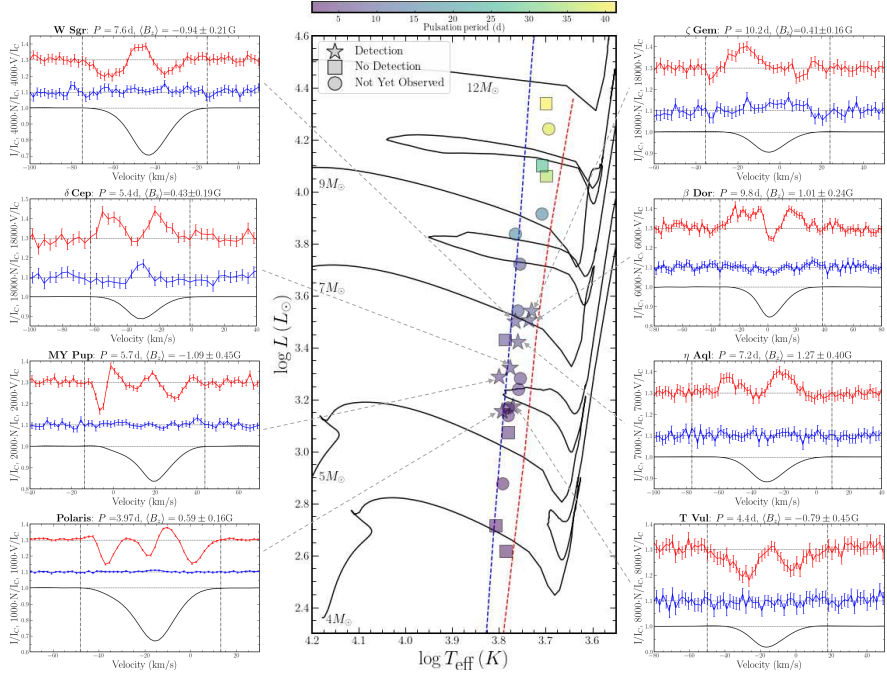}
 \caption{LSD Stokes $I$ (black), $V$ (red) and diagnostic null (blue) profiles for the Cepheid magnetic detections we have obtained to date. Stellar parameters, evolutionary tracks and IS boundaries from \cite{turner_2010},  \cite{luck_2018}, \cite{georgy_2013} and \cite{anderson_2016}.}
   \label{fig1}
\end{center}
\end{figure}

We have identified a magnitude-limited sample ($V<6$\,mag) of the brightest twenty-five Cepheids to perform a first systematic magnetic study (Fig.~\ref{fig1}). Before this survey, the only Cepheid with a confirmed magnetic detection was $\eta$~Aql \citep{grunhut_2010}. Initial results of this survey were reported by \cite{barron_2022}, including new magnetic detections of Polaris, $\delta$~Cep, and $\zeta$~Gem. Subsequently, we have obtained magnetic detections in four additional Cepheids: $\beta$~Dor, MY~Pup, T~Vul, and W~Sgr. 
\firstsection
\section{Results}
Fifteen survey targets have been observed to date with ESPaDOnS at the Canada-France-Hawaii Telescope and HARPSpol at the ESO 3.6m Telescope. We apply the multi-line procedure Least Squares Deconvolution (LSD) to the Stokes~$V$ spectra to diagnose the presence of a photospheric magnetic field. To detect weak magnetic signatures the Stokes~$V$ spectra are obtained at high signal-to-noise ratio ($\sim4000$ at 500 nm).

We detect Stokes~$V$ signatures in eight targets, demonstrating that magnetic fields are frequently detectable in Cepheids when observed with sufficient precision. Many targets show peculiar unipolar positive or negative Stokes~$V$ lobes, which are not predicted by standard Zeeman theory (e.g. $\delta$~Cep). These Stokes~$V$ profiles show similarities to those detected in some Am stars \citep{petit_2011}. We hypothesize that these features are due to the Zeeman effect modified by atmospheric velocity and magnetic field gradients. Polarized radiative transfer modelling of the Stokes~$V$ signatures could provide a new and unique probe of Cepheid atmospheric dynamics (see \citealt{barron_2022} and references therein).

In contrast, the LSD Stokes~$V$ profiles of Polaris and MY~Pup appear more similar to those observed in cool non-pulsating supergiants \citep{grunhut_2010}, with some net negative asymmetry. The lesser distortion in the Stokes~$V$ profiles may be related to the stars' low amplitude pulsations \citep{kienzle_1999, anderson_2019}. We have initiated a monitoring campaign of Polaris using ESPaDOnS to map its magnetic field using Zeeman-Doppler Imaging.
\firstsection
\bibliographystyle{apj}
\bibliography{./jbarron_iaus361}

\end{document}